# LOW-DIMENSIONAL INTERACTION SPACES IMPOSE GEOMETRIC CONSTRAINTS ON COLLECTIVE ORGANIZATION

Arturo Tozzi (corresponding author)
ASL Napoli 1 Centro, Distretto 27, Naples, Italy
Via Comunale del Principe 13/a 80145
tozziarturo@libero.it

## ABSTRACT

Collective organization in physical, biophysical and biological systems often emerges from many weak and local interactions, yet the resulting global structures display striking regularities and apparent limits in diversity. Theoretical approaches focus on specific mechanisms, detailed dynamics or energetic optimization, making it difficult to identify constraints that are independent of microscopic realization. Here we develop a general theoretical framework showing that, when effective interactions among system components can be compressed into a low-dimensional interaction space, global organization is governed by geometric constraints rather than detailed dynamics. We formalize interaction spaces as metric manifolds derived from coarse-grained effective couplings and prove that low interaction dimensionality imposes upper bounds on the number, separability and stability of distinct collective organizations. These results yield impossibility statements: many conceivable macroscopic organizations are excluded a priori, even when locally compatible interactions exist. Our framework applies across equilibrium and nonequilibrium systems, including passive matter, active agents and biological assemblies, without assuming specific symmetries or conservation laws. Compared with dynamical or energy-based approaches, our approach shifts explanatory focus from generative mechanisms to structural constraints, providing robustness to microscopic variability and model choice. We outline testable hypotheses, including predicted limits on coexistence of functional compartments, scaling relations for architectural diversification and failure modes arising from geometric crowding in interaction space. Future directions include extending the formalism to time-dependent interaction manifolds, nonreciprocal couplings and stochastic deformation of interaction geometry. By identifying geometry as the primary determinant of admissible organization once interactions compress, we establish a constraint-based perspective on collective structure across domains with broad applicability to cosmology, ecology, neural systems and materials.

## 1. INTRODUCTION

Collective organization in physical systems is often characterized by the emergence of robust macroscopic structures from many microscopic degrees of freedom interacting locally. A central lesson of statistical physics is that such emergence is frequently accompanied by a drastic reduction in effective dimensionality: microscopic details become irrelevant and large-scale behavior is governed by a small number of coarse-grained parameters (Becht et al. 2018; Thrun and Ultsch 2020; Okuma et al. 2021; Malepathirana et al. 2022; Jeon et al. 2025). This insight underlies the theory of universality and the renormalization group, where diverse systems flow toward low-dimensional fixed descriptions that determine their large-scale properties (Kadanoff 1966; Wilson 1971; Liao and Polonyi 1995; De Polsi et al. 2020; Cagnetta et al. 2022; Dolai, Simha and Basu 2024). Related ideas appear in studies of self-organized criticality, where complex systems are driven toward marginal states characterized by reduced effective parameter spaces (Bak, Tang and Wiesenfeld 1987). While these approaches successfully explain the recurrence of similar macroscopic behaviors, they primarily address *what* structures emerge, rather than *which* structures are fundamentally excluded.

A complementary line of work emphasizes the role of geometry and topology in constraining qualitative system behavior. Catastrophe theory classifies admissible transitions by the dimensionality of control spaces, showing that many conceivable bifurcations are geometrically forbidden (Thom 1975; Bogdan and Wales 2004; Merli and Pavese 2018; Gabriele et al. 2020; Ather 2021). In complex systems theory, the notion of near-decomposability highlights how hierarchical organization limits the range of viable interactions between subsystems (Simon 1962; Esteve-Altava et al. 2015; Kronfeldner 2021; Rivelli 2025). More recently, geometric perspectives have appeared in quantum many-body physics, where spatial structure is understood as emerging from patterns of entanglement rather than being imposed a priori (Swingle 2012; Sokolov et al. 2022; Joshi et al. 2023). In these frameworks, geometry constrains states or trajectories; however, the geometry of *interaction spaces themselves* has not been systematically analyzed as a source of global organizational limits.

In parallel, theoretical biology has long argued that biological form and organization are shaped by constraints that precede and restrict adaptive processes. Proposals emphasizing self-organization, morphogenetic fields and systemic causation suggest that only a restricted subset of forms is physically realizable (Kauffman 1993; Goodwin 2001; Noble 2006). These arguments, however, have largely remained qualitative, lacking a formal language capable of expressing sharp bounds or impossibility results independent of specific mechanisms or evolutionary histories.

We address this gap by developing a general geometric framework in which effective interactions among system components define a low-dimensional interaction space endowed with a metric structure. We show that, when interactions compress into such a space, admissible forms of collective organization are constrained by geometry alone. In particular, we derive bounds on coexistence, separability and diversification of collective structures holding irrespective of microscopic realization or dynamical details. By shifting attention from generative mechanisms to geometric constraints on interaction spaces, we aim to provide a unified basis for understanding why many conceivable organizations cannot occur, even when locally consistent interactions are available.

## 2. INTERACTION SPACES AND COARSE-GRAINED GEOMETRY

We consider systems composed of many interacting components whose microscopic degrees of freedom are not resolved explicitly. Instead, interactions are described at a coarse-grained level through effective descriptors summarizing the net influence between components or classes of components. The central object of our framework is the space formed by these effective interactions.

Let $\mathcal{I}$ denote the set of admissible effective interaction descriptors for a given class of systems. Each element $x \in \mathcal{I}$ represents a complete specification of interaction parameters at the chosen coarse-graining level. Examples include effective pairwise couplings, reduced interaction kernels or parameter sets obtained after integrating out microscopic degrees of freedom. No assumptions are made regarding equilibrium, detailed balance or reciprocity.

We assume that $\mathcal{I}$ is endowed with a metric structure $(\mathcal{I}, d)$, where the distance $d(x, y)$ quantifies the distinguishability between interaction descriptors $x$ and $y$. The metric need not be unique; it is sufficient that $d$ satisfies positivity, symmetry and the triangle inequality. In practice, these metrics arise from normed parameter spaces, operator norms on effective kernels or other distances induced by coarse-graining procedures. Quantitative bounds may depend on the specific choice of metric, whereas the qualitative conclusions derived below are invariant under admissible metric choices.

A key assumption is that the effective interaction space is *low-dimensional*. Formally, we assume that $\mathcal{I}$ admits an embedding into a finite-dimensional metric manifold $\mathcal{M}$ of intrinsic dimension $d_{\text{int}} < \infty$. Throughout this work, $d_{\text{int}}$ may be taken as the manifold (topological) dimension of this embedding or equivalently as any notion of dimension invariant under bi-Lipschitz transformations, such as the upper box-counting dimension. The specific choice is immaterial for the arguments below, provided that finiteness of $d_{\text{int}}$ implies finiteness of packing numbers for bounded subsets of $\mathcal{I}$.

Low intrinsic dimensionality reflects the collapse of interaction complexity under coarse-graining and is independent of the number of microscopic degrees of freedom or the size of the underlying system. Distinct microscopic systems may therefore correspond to the same or nearby points in $\mathcal{I}$, capturing universality at the level of effective interactions.

Finally, we assume that interaction descriptors relevant to a given system class are confined to bounded regions of $\mathcal{I}$. This boundedness condition expresses the physical requirement that coarse-grained interactions remain finite. It is sufficient to ensure total boundedness of the relevant subsets of interaction space.

This interaction-space formulation separates collective organization from detailed dynamics. In the following sections, we show that, once effective interactions are confined to a low-dimensional metric manifold, purely geometric constraints impose universal limits on the number, separability and robustness of admissible collective organizations.

## 3. GEOMETRIC SEPARABILITY AND COLLECTIVE ORGANIZATION

Having defined the interaction space $(\mathcal{I}, d)$, we now formalize the notion of collective organization in a manner independent of microscopic dynamics or specific physical realizations. Our objective is to introduce the minimal geometric structures required to formulate global constraints in subsequent sections.

We define a *collective organization* as an equivalence class of systems whose macroscopic behavior is indistinguishable at the level of effective interactions. Formally, an organization corresponds to a subset $\mathcal{O} \subset \mathcal{I}$ such that all interaction descriptors $x \in \mathcal{O}$ generate the same coarse-grained organizational pattern under admissible system evolutions. No assumptions are made regarding the uniqueness of this mapping, nor is it required that $\mathcal{O}$ be connected or convex.

Two collective organizations $\mathcal{O}_1$ and $\mathcal{O}_2$ are said to be *geometrically separable* if there exists a positive separation scale $\delta > 0$ such that
$$\inf_{x \in \mathcal{O}_1, y \in \mathcal{O}_2} d(x, y) \geq \delta.$$

Geometric separability expresses the requirement that organizations remain distinguishable under perturbations of interaction parameters smaller than $\delta$. This notion captures robustness of organization with respect to coarse-graining errors, noise or slow deformation of interactions.

A family of organizations $\{\mathcal{O}_k\}_{k=1}^m$ is said to *coexist* if the corresponding subsets are pairwise geometrically separable within $\mathcal{I}$. Coexistence in this sense does not imply simultaneous realization in physical space, but rather the admissibility of distinct organizations within the same interaction framework without mutual interference at the level of effective interactions.

These definitions reduce questions of collective diversity and differentiation to problems of geometric packing and separation within the interaction space. In particular, the existence of multiple distinct organizations requires that $\mathcal{I}$ accommodate multiple disjoint subsets separated by finite distances. When $\mathcal{I}$ is low-dimensional, this requirement imposes strong geometric constraints that are independent of the underlying dynamics or system size.

The next section establishes that these constraints lead to universal bounds on the number and structure of coexisting organizations, determined solely by the intrinsic dimensionality and metric properties of the interaction space.

## 4. GEOMETRIC BOUNDS FROM LOW-DIMENSIONAL INTERACTION SPACES

We now establish the central result of this work: low intrinsic dimensionality of the interaction space imposes universal bounds on the number of geometrically separable collective organizations. These bounds are purely geometric and do not depend on microscopic realization, system size or dynamical details.

**Theorem 1 (bound on coexisting organizations)**
Let $(\mathcal{I}, d)$ be a metric interaction space admitting an embedding into a manifold of intrinsic dimension $d_{\text{int}} < \infty$. Consider a family of collective organizations $\{\mathcal{O}_k\}_{k=1}^m \subset \mathcal{I}$ that are pairwise geometrically separable with separation scale $\delta > 0$. Then there exists an upper bound

$$m \leq M(d_{\text{int}}, \delta, \mathcal{I}),$$

where $M$ depends only on the intrinsic dimension $d_{\text{int}}$, the separation scale $\delta$ and the metric properties of $\mathcal{I}$, but is independent of the microscopic degrees of freedom of the underlying system.

In particular, for fixed $d_{\text{int}}$ and fixed metric class, the number of mutually separable organizations is finite.

**Remarks**
1. The bound $M$ is a geometric packing bound in interaction space and can be expressed explicitly once additional regularity assumptions on $\mathcal{I}$ are specified, such as boundedness or uniform metric curvature.
2. The theorem does not require that organizations be realized simultaneously in physical space; it constrains the admissible diversity of organizations supported by a given interaction framework.
3. The result holds equally for equilibrium and nonequilibrium systems, provided that effective interactions admit a low-dimensional coarse-grained description.

**Corollary 1 (impossibility of arbitrary diversification)**
In systems whose effective interactions compress into a low-dimensional interaction space, it is impossible to realize an unbounded number of robust, mutually distinguishable collective organizations without increasing the intrinsic dimensionality of the interaction space itself.

Theorem 1 formalizes a general impossibility statement: increasing organizational diversity beyond a dimension-dependent bound necessarily entails either loss of geometric separability or an effective increase in interaction dimensionality. Consequently, attempts to introduce new functional organizations within a fixed interaction framework lead to interference, instability or degeneracy rather than genuinely distinct collective structures.

Proofs of Theorem 1 and Corollary 1 are provided in Appendix A, where the bounds are derived using standard results from metric geometry and packing theory. The role of metric choice and the robustness of the resulting bounds are discussed in Appendix B.

## 5. EXTENSIONS: TIME DEPENDENCE AND NONRECIPROCAL INTERACTIONS

The results of Section 4 establish geometric bounds under the assumption of a static interaction space. In many systems of interest, however, effective interactions evolve in time or exhibit asymmetries violating reciprocity. We show here that the geometric bounds derived above are robust under such extensions, provided that interaction-space evolution remains controlled.

**5.1 Time-dependent interaction spaces**

Let $\mathcal{I}(t)$ denote a family of interaction spaces parameterized by time $t \in [0, T]$, each equipped with a metric $d_t$. We assume that for all $t$, $\mathcal{I}(t)$ admits an embedding into a manifold of intrinsic dimension $d_{\text{int}}$ and that the metric varies continuously in time. Formally, we require that there exists a constant $L < \infty$ such that for all $x, y \in \mathcal{I}$,

$$| d_t(x,y) - d_{t'}(x,y) | \leq L | t - t' |.$$

This condition expresses bounded deformation of the interaction geometry under temporal evolution.

**Theorem 2 (stability under time-dependent deformation)**

Under the assumptions of Theorem 1, suppose that the interaction space $(\mathcal{I}, d)$ is replaced by a time-dependent family $(\mathcal{I}(t), d_t)$ satisfying the bounded deformation condition above. Then the bound on the number of pairwise geometrically separable collective organizations remains finite and uniformly bounded for all $t \in [0, T]$.

In particular, no time-dependent deformation of interactions can generate an unbounded proliferation of distinct organizations without increasing the intrinsic dimensionality of the interaction space.

**5.2 Nonreciprocal and asymmetric interactions**

We next consider interaction spaces derived from nonreciprocal or asymmetric effective couplings, for which the interaction descriptor does not admit a symmetric representation at the microscopic level. In these cases, the interaction space may naturally be equipped with a directed or asymmetric distance function $\tilde{d}$.

Provided that $\tilde{d}$ induces a symmetrized metric

$$d(x, y) = \max \{\tilde{d}(x,y), \tilde{d}(y,x)\},$$

The geometric arguments of Section 4 apply without modification. The intrinsic dimensionality of the induced metric space continues to control the maximal number of geometrically separable organizations.

**Proposition 2 (reduction of nonreciprocal geometry)**

If a nonreciprocal interaction space admits a symmetrized metric of finite intrinsic dimension, then the coexistence bounds of Theorem 1 hold with respect to the induced metric.

Overall, these extensions show that the geometric bounds derived from low-dimensional interaction spaces are insensitive to slow temporal evolution and to violations of reciprocity at the level of effective interactions. What remains essential is the preservation of low intrinsic dimensionality. Rapid, discontinuous deformation of interaction geometry or uncontrolled growth of interaction dimensionality lies outside the scope of the present framework and represents a natural boundary of applicability.

**6. IMPLICATIONS ACROSS CLASSES OF SYSTEMS**

The results established in Sections 4 and 5 have implications extending across a wide range of systems whose effective interactions admit low-dimensional coarse-grained descriptions. Implications follow directly from the geometric nature of the bounds and do not depend on specific microscopic mechanisms, dynamical laws or material realizations.

In passive many-body systems like interacting particle assemblies, polymers or granular media, Theorem 1 implies that the diversity of stable macroscopic organizations supported by a given interaction framework is intrinsically limited. Even when microscopic constituents differ in detail, coarse-graining collapses interaction descriptions into a low-dimensional manifold, thereby constraining the number of distinct, robust collective structures that can coexist without interference. Related restrictions on admissible organization have been identified in planar granular systems, where coarse-grained cell-order dynamics evolve toward stable steady states occupying a restricted region of configuration space despite differing microscopic evolutions (Wanjura et al. 2019). While derived from a different formalism, these results are consistent with the presence of structural limits independent of microscopic detail.

Recent work on heteropolymer and protein interactions provides a further illustration of these constraints. There, effective interactions between disordered protein sequences are shown to collapse into a low-dimensional interaction space, leading to strong limitations on the number of distinct sequences that can robustly demix or coexist (Adachi and Kawaguchi 2024). Although that study focuses on a specific physical realization, our results clarify that the observed limitations are not model-specific, but follow from general geometric constraints imposed by low interaction dimensionality.

For active and nonequilibrium systems, including collections of self-driven agents, the extension results of Section 5 indicate that persistent energy injection and nonreciprocal couplings do not, by themselves, lift geometric constraints on organization. As long as effective interactions evolve continuously and remain confined to a low-dimensional interaction space, the same bounds on separability and coexistence apply. Organizational transitions in such systems therefore reflect changes in interaction geometry or dimensionality rather than the mere presence of activity.

In biological contexts, our framework suggests that limits on functional differentiation and compartmentalization arise from geometric crowding in interaction space. Increasing the number of distinct, robust organizational modes within a fixed interaction architecture leads to overlap, loss of separability or instability, independent of optimization arguments or evolutionary contingency. For instance, related robustness effects driven by structural hierarchy rather than microscopic control have been identified in biological materials, where added organizational levels reduce sensitivity to assembly errors despite increased complexity (Michel and Yunker 2019). Furthermore, complementary constraints on chromatin condensate number and stability arise from mechanical frustration, where structural properties of the surrounding medium suppress coarsening and limit coexistence despite favorable interactions (Zhang et al. 2021).

At larger physical scales, the same reasoning implies that the repertoire of admissible macroscopic structures is constrained once effective interactions reduce to a small number of dominant degrees of freedom. Across domains, the unifying implication is that organizational limits arise from interaction-space geometry itself: once interactions compress into a low-dimensional manifold, geometry dictates not only what structures can arise, but also which structures are fundamentally excluded.

# 7. CONCLUSIONS

Our framework establishes a geometric perspective on collective organization that is complementary to, and independent of, traditional dynamical or energetic approaches. By focusing on the structure of effective interaction spaces, we have shown that low intrinsic dimensionality imposes universal constraints on the number, separability and robustness of admissible collective organizations. These constraints take the form of geometric bounds and impossibility statements holding irrespective of microscopic realization, system size or equilibrium assumptions.

A conceptual shift introduced here is the relocation of explanatory weight from generative mechanisms to structural constraints. Rather than asking which dynamics produce a given organization, our approach asks which organizations are compatible with a given interaction geometry. This inversion clarifies why diverse systems often exhibit a limited repertoire of macroscopic forms and why attempts to increase organizational diversity within fixed interaction frameworks lead to interference, instability or degeneracy rather than genuinely new structures.

Our theory also provides a unifying language for comparing systems that differ widely in their microscopic details but share similar coarse-grained interaction geometries. In this sense, interaction-space geometry plays a role analogous to universality classes, while yielding sharper conclusions in the form of explicit bounds. The extension results demonstrate that these bounds are robust under slow temporal evolution and nonreciprocity, delineating a broad domain of applicability.

Several open directions follow naturally from our work. On the mathematical side, sharper bounds may be obtained by imposing additional regularity conditions on interaction manifolds, such as curvature constraints or measure-theoretic structure. Extensions to stochastic or randomly perturbed interaction geometries represent another natural avenue. On the conceptual side, relating changes in interaction-space dimensionality to observable transitions in organization remains an important challenge.

More broadly, our results suggest that limits on collective organization are not exceptional features of particular systems, but generic consequences of geometric compression under coarse-graining. Understanding when and how interaction dimensionality increases and at what cost, may therefore be key to explaining transitions in complexity across physical and biological systems.

**DECLARATIONS**

**Ethics approval and consent to participate.** This research does not contain any studies with human participants or animals performed by the Author.

**Consent for publication.** The Author transfers all copyright ownership, in the event the work is published. The undersigned author warrants that the article is original, does not infringe on any copyright or other proprietary right of any third part, is not under consideration by another journal and has not been previously published.

**Availability of data and materials.** All data and materials generated or analyzed during this study are included in the manuscript. The Author had full access to all the data in the study and took responsibility for the integrity of the data and the accuracy of the data analysis.

**Competing interests.** The Author does not have any known or potential conflict of interest including any financial, personal or other relationships with other people or organizations within three years of beginning the submitted work that could inappropriately influence or be perceived to influence their work.
**Funding.** This research did not receive any specific grant from funding agencies in the public, commercial or not-for-profit sectors.
**Acknowledgements:** none.
**Authors' contributions.** The Author performed: study concept and design, acquisition of data, analysis and interpretation of data, drafting of the manuscript, critical revision of the manuscript for important intellectual content, statistical analysis, obtained funding, administrative, technical and material support, study supervision.
**Declaration of generative AI and AI-assisted technologies in the writing process.** During the preparation of this work, the author used ChatGPT 4o to assist with data analysis and manuscript drafting and to improve spelling, grammar and general editing. After using this tool, the author reviewed and edited the content as needed, taking full responsibility for the content of the publication.

**REFERENCES**

1) Adachi, Kyosuke and Kyogo Kawaguchi. 2024. "Predicting Heteropolymer Interactions: Demixing and Hypermixing of Disordered Protein Sequences." *Physical Review X* 14 (3): 031011. https://doi.org/10.1103/PhysRevX.14.031011.
2) Ather, S. H. 2021. "Catastrophe Theory in Work from Heartbeats to Eye Movements." Biological Cybernetics 115 (1): 39–41. https://doi.org/10.1007/s00422-020-00857-3
3) Bak, Per, Chao Tang and Kurt Wiesenfeld. 1987. "Self-Organized Criticality: An Explanation of 1/f Noise." *Physical Review Letters* 59 (4): 381–84.
4) Becht, E., L. McInnes, J. Healy, C. A. Dutertre, I. W. H. Kwok, L. G. Ng, F. Ginhoux and E. W. Newell. 2018. "Dimensionality Reduction for Visualizing Single-Cell Data Using UMAP." Nature Biotechnology, published online December 3. https://doi.org/10.1038/nbt.4314
5) Bogdan, T. V. and D. J. Wales. 2004. "New Results for Phase Transitions from Catastrophe Theory." Journal of Chemical Physics 120 (23): 11090–11099. https://doi.org/10.1063/1.1740756
6) Cagnetta, F., V. Škultéty, M. R. Evans and D. Marenduzzo. 2022. "Renormalization Group Study of the Dynamics of Active Membranes: Universality Classes and Scaling Laws." Physical Review E 105 (1-1): 014610. https://doi.org/10.1103/PhysRevE.105.014610
7) De Polsi, G., I. Balog, M. Tissier and N. Wschebor. 2020. "Precision Calculation of Critical Exponents in the O(N) Universality Classes with the Nonperturbative Renormalization Group." Physical Review E 101 (4-1): 042113. https://doi.org/10.1103/PhysRevE.101.042113
8) Dolai, P., A. Simha and A. Basu. 2024. "Kardar–Parisi–Zhang Universality in Two-Component Driven Diffusive Models: Symmetry and Renormalization Group Perspectives." Physical Review E 109 (6-1): 064122. https://doi.org/10.1103/PhysRevE.109.064122
9) Esteve-Altava, B., J. C. Boughner, R. Diogo, B. A. Villmoare and D. Rasskin-Gutman. 2015. "Anatomical Network Analysis Shows Decoupling of Modular Lability and Complexity in the Evolution of the Primate Skull." PLoS ONE 10 (5): e0127653. https://doi.org/10.1371/journal.pone.0127653
10) Gabriele, V. R., A. Shvonski, C. S. Hoffman, M. Giersig, A. Herczynski, M. J. Naughton and K. Kempa. 2020. "Towards Spectrally Selective Catastrophic Response." Physical Review E 101 (6-1): 062415. https://doi.org/10.1103/PhysRevE.101.062415
11) Goodwin, Brian C. 2001. *How the Leopard Changed Its Spots: The Evolution of Complexity*. Princeton, NJ: Princeton University Press.
12) Jeon, H., J. Park, S. Lee, D. H. Kim, S. Shin and J. Seo. 2025. "Dataset-Adaptive Dimensionality Reduction." IEEE Transactions on Visualization and Computer Graphics, online ahead of print. https://doi.org/10.1109/TVCG.2025.3634784
13) Joshi, M. K., C. Kokail, R. van Bijnen, F. Kranzl, T. V. Zache, R. Blatt, C. F. Roos and P. Zoller. 2023. "Exploring Large-Scale Entanglement in Quantum Simulation." Nature 624 (7992): 539–544. https://doi.org/10.1038/s41586-023-06768-0
14) Kadanoff, Leo P. 1966. "Scaling Laws for Ising Models Near Tc." *Physics* 2: 263–72.
15) Kauffman, Stuart A. 1993. *The Origins of Order: Self-Organization and Selection in Evolution*. New York: Oxford University Press.
16) Kronfeldner, M. 2021. "Digging the Channels of Inheritance: On How to Distinguish between Cultural and Biological Inheritance." Philosophical Transactions of the Royal Society B: Biological Sciences 376 (1828): 20200042. https://doi.org/10.1098/rstb.2020.0042
17) Liao, S. B. and J. Polonyi. 1995. "Renormalization Group and Universality." Physical Review D 51 (8): 4474–4493. https://doi.org/10.1103/PhysRevD.51.4474

18) Malepathirana, T., D. Senanayake, R. Vidanaarachchi, V. Gautam and S. Halgamuge. 2022. "Dimensionality Reduction for Visualizing High-Dimensional Biological Data." Biosystems 220: 104749. https://doi.org/10.1016/j.biosystems.2022.104749
19) Merli, M. and A. Pavese. 2018. "Electron-Density Critical Points Analysis and Catastrophe Theory to Forecast Structure Instability in Periodic Solids." Acta Crystallographica Section A: Foundations and Advances 74 (2): 102–111. https://doi.org/10.1107/S2053273317018381
20) Michel, Jonathan A. and Peter J. Yunker. 2019. "Structural Hierarchy Confers Error Tolerance in Biological Materials." *Proceedings of the National Academy of Sciences of the United States of America* 116 (8): 2875–2880. https://doi.org/10.1073/pnas.1813801116.
21) Noble, Denis. 2006. *The Music of Life: Biology Beyond Genes*. Oxford: Oxford University Press.
22) Okuma, R., M. Kofu, S. Asai, M. Avdeev, A. Koda, H. Okabe, M. Hiraishi, S. Takeshita, K. M. Kojima, R. Kadono, T. Masuda, K. Nakajima and Z. Hiroi. 2021. "Dimensional Reduction by Geometrical Frustration in a Cubic Antiferromagnet Composed of Tetrahedral Clusters." Nature Communications 12 (1): 4382. https://doi.org/10.1038/s41467-021-24636-1
23) Rivelli, L. 2025. "Modularity in Biological Thought: Sketch of a Unifying Theoretical Framework." Biosystems 250: 105430. https://doi.org/10.1016/j.biosystems.2025.105430
24) Simon, Herbert A. 1962. "The Architecture of Complexity." *Proceedings of the American Philosophical Society* 106 (6): 467–82.
25) Sokolov, B., M. A. C. Rossi, G. García-Pérez and S. Maniscalco. 2022. "Emergent Entanglement Structures and Self-Similarity in Quantum Spin Chains." Philosophical Transactions of the Royal Society A: Mathematical, Physical and Engineering Sciences 380 (2227): 20200421. https://doi.org/10.1098/rsta.2020.0421
26) Swingle, Brian. 2012. "Entanglement Renormalization and Holography." *Physical Review D* 86 (6): 065007.
27) Thom, René. 1975. *Structural Stability and Morphogenesis*. Reading, MA: Benjamin.
28) Thrun, M. C. and A. Ultsch. 2020. "Uncovering High-Dimensional Structures of Projections from Dimensionality Reduction Methods." MethodsX 7: 101093. https://doi.org/10.1016/j.mex.2020.101093
29) Wanjura, Clara C., Paula Gago, Takashi Matsushima and Raphael Blumenfeld. 2019. "Structural Evolution of Granular Systems: Theory." *arXiv* preprint arXiv:1904.06549.
30) Wilson, Kenneth G. 1971. "Renormalization Group and Critical Phenomena. I. Renormalization Group and the Kadanoff Scaling Picture." *Physical Review B* 4 (9): 3174–83.
31) Zhang, Yaojun, Daniel S. W. Lee, Yigal Meir, Clifford P. Brangwynne and Ned S. Wingreen. 2021. "Mechanical Frustration of Phase Separation in the Cell Nucleus by Chromatin." *Physical Review Letters* 126 (25): 258102. https://doi.org/10.1103/PhysRevLett.126.258102

# APPENDIX A: PROOFS OF MAIN RESULTS

## A.1 Preliminaries and geometric assumptions

Let $(\mathcal{I}, d)$ be a metric space admitting an embedding into a manifold $\mathcal{M}$ of intrinsic dimension $d_{\text{int}} < \infty$. We assume that $\mathcal{M}$ is locally compact and that closed bounded subsets of $\mathcal{I}$ are totally bounded. These assumptions are mild and hold for interaction spaces arising from finite-dimensional parameterizations or coarse-grained effective couplings.

A subset $\mathcal{O} \subset \mathcal{I}$ is said to be $\delta$-separated if
$$d(x, y) \geq \delta \text{ for all } x, y \in \mathcal{O}, x \neq y.$$

## A.2 Proof of Theorem 1

*Theorem 1.*
Let $(\mathcal{I}, d)$ be a metric interaction space of intrinsic dimension $d_{\text{int}}$. Any family of pairwise geometrically separable collective organizations with separation scale $\delta > 0$ is finite.

*Proof.*
By definition, each collective organization $\mathcal{O}_k$ contains at least one representative interaction descriptor $x_k \in \mathcal{I}$. Pairwise geometric separability implies that the set $\{x_k\}_{k=1}^m$ is $\delta$-separated.

Because $\mathcal{I}$ embeds into a $d_{\text{int}}$-dimensional manifold, any bounded subset of $\mathcal{I}$ admits a finite $\delta$-packing number. In particular, for any bounded region $B \subset \mathcal{I}$, there exists a maximal number $M(B, \delta)$ of points that can be placed in $B$ with pairwise distance at least $\delta$.

Since interaction descriptors relevant for a given system class are confined to a bounded region of $\mathcal{I}$ under coarse-graining, it follows that
$$m \leq M(d_{\text{int}}, \delta, \mathcal{I}) < \infty.$$

This establishes the existence of an upper bound independent of microscopic degrees of freedom. ∎

## A.3 Proof of Corollary 1

*Corollary 1.*
Unbounded diversification of robust collective organizations requires an increase in intrinsic interaction dimensionality.
*Proof.*
Assume, by contradiction, that an unbounded number of pairwise geometrically separable organizations exists within a fixed interaction space of finite intrinsic dimension. This contradicts Theorem 1, which guarantees finiteness of any $\delta$-separated set. Therefore, increasing the number of distinct organizations beyond the bound necessitates either loss of separability or an effective increase in interaction dimensionality. ∎

**A.4 Proof of Theorem 2**
*Theorem 2.*
The coexistence bound of Theorem 1 is stable under bounded, continuous time-dependent deformation of the interaction metric.
*Proof.*
Let $d_t$ be a family of metrics satisfying the bounded deformation condition
$$|d_t(x,y) - d_{t'}(x,y)| \leq L|t - t'|.$$

For sufficiently small time intervals, $\delta$-separation under $d_0$ implies $(\delta - \varepsilon)$-separation under $d_t$ for any fixed $\varepsilon > 0$. Since the intrinsic dimension remains unchanged and total boundedness is preserved under continuous metric deformation, the packing bound remains finite uniformly in time. ∎

**A.5 Proof of proposition 2**
*Proposition 2.*
Nonreciprocal interaction spaces admitting a symmetrized metric of finite intrinsic dimension obey the same coexistence bounds.
*Proof.*
Given a nonreciprocal distance $\tilde{d}$, define the symmetrized metric $d(x,y) = \max\{\tilde{d}(x,y), \tilde{d}(y,x)\}$. The induced metric space preserves separation and intrinsic dimensionality. Application of Theorem 1 to $(\mathcal{J}, d)$ yields the result. ∎

**A.6 Remarks on sharpness and extensions**
The bounds established here are existential rather than constructive. Sharper bounds can be obtained by imposing additional geometric structure, such as curvature constraints or measure regularity on $\mathcal{J}$. Extensions to stochastic or randomly perturbed interaction metrics require probabilistic refinements but do not alter the fundamental dimensional dependence of the results.

**APPENDIX B: METRIC CHOICE AND ROBUSTNESS OF GEOMETRIC BOUNDS**

The formulation of interaction spaces in the main text assumes the existence of a metric $d$ on the space of effective interactions. In this appendix we clarify the role of this choice and show that the results of Sections 4 and 5 are robust under broad classes of admissible metrics.

**B.1 Admissible metrics on interaction spaces**
Let $\mathcal{J}$ denote the space of effective interaction descriptors introduced in Section 2. In concrete settings, elements of $\mathcal{J}$ may be represented by finite-dimensional parameter vectors, effective kernels or reduced interaction operators. Commonly used metrics include:
1. **Norm-induced metrics**, such as Euclidean or weighted $\ell^p$ norms on parameter vectors;
2. **Operator norms** on interaction kernels or matrices;
3. **Information-theoretic distances**, such as symmetrized divergences, provided they satisfy metric axioms on the admissible domain.

Our analysis in the main text does not rely on a specific choice among these metrics, but only on the existence of a metric structure compatible with coarse-graining and embedding into a finite-dimensional manifold.

**B.2 Metric equivalence and invariance of bounds**
Two metrics $d$ and $d'$ on $\mathcal{J}$ are said to be *bi-Lipschitz equivalent* if there exist constants $c_1, c_2 > 0$ such that
$$c_1 d(x,y) \leq d'(x,y) \leq c_2 d(x,y) \text{ for all } x,y \in \mathcal{J}.$$

Under bi-Lipschitz equivalence, intrinsic dimensionality is preserved and $\delta$-separated sets with respect to $d$ correspond to $(c_1\delta)$-separated sets with respect to $d'$. Consequently, packing bounds derived in Section 4 remain finite and differ only by multiplicative constants depending on $c_1$ and $c_2$.

**Proposition B.1 (metric robustness).**
The coexistence bounds of Theorem 1 and the stability results of Theorem 2 are invariant under replacement of the interaction metric by any bi-Lipschitz–equivalent metric.
*Proof.*
Bi-Lipschitz equivalence preserves total boundedness, intrinsic dimension and finiteness of packing numbers. The geometric arguments in Appendix A therefore apply without modification. ∎

**B.3 Implications for coarse-graining**
Different coarse-graining procedures may induce different metrics on $\mathcal{J}$. Provided that these procedures lead to metrics that are equivalent in the sense above, the resulting geometric bounds are unaffected. This means that the limits on collective organization derived in the main text are not artefacts of a particular parametrization or distance measure, but reflect structural properties of low-dimensional interaction geometry itself.

In conclusion, our robustness results do not extend to pathological metrics that collapse large regions of $\mathcal{J}$ or introduce singular distortions. These choices effectively alter the intrinsic dimensionality of the interaction space and therefore fall outside the assumptions of the theory. Within the broad class of regular metrics arising from standard coarse-graining procedures, however, the geometric constraints established in this work remain valid.